\documentclass[12pt]{article}
 \usepackage[dvips]{graphicx}
\begin{document}
 \setcounter{section}{0}
\title{Scale Relativity in Cantorian  $\cal E^{(\infty)}$ Space and
 Average Dimensions of Our World}
\author{Carlos Castro\thanks{Center for Theoretical Studies of
                             Physical Systems Clark Atlanta University,
                             Atlanta, GA. 30314}~  ~Alex Granik\thanks
                             {Department of Physics
                     University of the Pacific, Stockton, CA. 95211}~~ M.S.El Naschie }
 \date{}
 \maketitle

 \begin{abstract}

Cantorian fractal spacetime, a family member of  von Neumann's
noncommutative geometry, is introduced as a geometry underlying a
new relativity theory which is similar to the relation between
general relativity and Riemannian geometry. Based on this model
and the new relativity theory an ensemble distribution of all the
dimensions of quantum spacetime is derived with the help of Fermat
last theorem. The calculated average dimension is very close to
the value of $4$+$\phi^3 $ (where $\phi$ is the golden mean)
obtained by El Naschie on the basis of a  different approach. It
is shown that within the framework of the  new relativity  the
cosmological constant problem is nonexistent,  since the Universe
self-organizes and self-tunes according to the  renormalization
group (RG) flow with respect to a local scaling  microscopic
arrow of time. This implies that the world emerged as  a result
of a non-equilibrium process of self-organized critical phenomena
launched by vacuum fluctuations in Cantorian fractal spacetime
$\cal E^{\infty}$. It is shown that we are  living in a
metastable vacuum  and are moving towards a fixed point ( $
D_{av}$ = $4$ + $\phi^3$) of the RG. After reaching this point, a
new phase transition will drive the universe to a quasi-crystal
phase of the lower average dimension  of $\phi^3$. \
\end{abstract}
\pagebreak

\section{Introduction}

Rephrasing I.Kant we can say that any physical theory should start
from the first principles and assumptions based on intuition
directly related to the existing physical phenomena, then proceed
to the concepts, complete theory with a central  idea , and crown
it with the experimental verification and testable predictions.
One of the best examples of this path is provided by the theory of
relativity and quantum mechanics. Their further development
spurred by the need to unite gravity and quantum mechanics
resulted in the birth of the string theory.

During 30 years of its development the string theory demonstrated
remarkable achievements in all but one respect: there is no direct
way to verify its theoretical results. In addition, there is at
least one "thorny" question: why does the world we are living in
appears to be $3+1$-dimensional? Up to now  string theory could not
produced a satisfactory answer. It seems that string theory
ignored possible alternative basic assumptions which, as we will
see later, could have produced a plausible answer:
multi-fractality of spacetime, a possibility of spacetime
dimensions ranging from negative dimensions to infinite ones,as
in the $\cal E ^\infty$ Cantorian theory of El Naschie \cite{en2}
and the resolution-dependent character of physical laws as
described for example by Nottale's scale relativity \cite{ln1}.

One of the basic  problem facing future developments of a
successful physical theory is a description of the quantum
"reality" without introducing by hand  {\it a priori} existing
background. Interestingly enough, as one of the first steps in
formulating the theoretical foundations of physics Newton
disposed of this problem rather easily by simply introducing
(almost at the very beginning of "Principia") the absolute space
and time as entities "which do not bear any relation to anything
external". It is clear that without an attendant background (
either not changing or changing rather slowly as compared to
developing phenomena) any description of a physical theory looks
impossible.

To formalize dynamics of physical phenomena we have to answer the
following question: the dynamics with  respect to what? For
example, the Lagrangian of the string theory refers to the
background of proper time (which is rather a vague notion since
we have to have a physical device attached to a string and
measuring time intervals which in turn are not defined
precisely) and coordinates along the string. The very spacetime
exists only to "the extent that it can be reconstructed from the
2-D field theory"\cite{ew1}.

 n other words, there is no " background spacetime " $ab ~
initio$. Quantum Spacetime is truly a " process in the making "
\cite{aa1} which implies that gravity should be described by a
{\bf nonlinear} theory. We are dealing with a nonlinear complex
dynamical system that is able of self-tuning in a fashion  similar
to biological systems in nature \cite{ls1}. To put it succinctly,
we can say that the universe could be viewed as the  "organism",
or using a more fashionable wording, the ultimate  quantum Machian
computer. Here we refer to the Mach principle: a physical theory
should contain only the {\it relations} between physical
quantities. It is the relations (reflected in algebraic
operations) among their representative elements ( monads) that
govern a system's evolution \cite{rc1,tn1}. Therefore
algebraic-categorical relations are the only meaningful relations
which are applicable for a description of such an "organism"
\cite{cc1}- \cite{ls2}.

 he problem of spacetime generation could be addressed with the
help of what we call a new relativity
\cite{cc1}-\cite{cc4},\cite{cc5}, a theory which is based on a few
postulates. One of these postulates is an old bootstrap idea by
Chew: a physical system in a process of its development
simultaneously generates its own background. In particular,
geometry ( initially nonexistent) is produced by a recursive (
self-referential) process starting from a "hyperpoint"  \cite{cc4}
endowed with infinite dimensions, or equivalently with an
infinite amount of information \cite{en2},\cite{en1}-\cite{go1}.

The latter implies a possibility of emergence of an ensemble of
interacting universes having all possible dimensions  ranging
from negative values to infinity. The ensemble average yields the
3-dimensional world of our perception. Thus an assumption about
existence of such an ensemble ( emerging as  a result of a
bootstrapping process) of the universes with all possible
dimensions points in the direction where even string theory with
its 10, 11, 26 dimensions does not go. Implications of infinite
dimensions which in the final run can be viewed as "words"
constructed out of infinite information contained in a
"hyperpoint" provides us with a source of a bootstrapping
mechanism of  spacetime generation. Equally, the energy also
emerges as a result of self-organization of the initial infinite
amount of information (number of bits).

The inter-connectivity of energy, dimensions, information becomes
especially visible if we consider the information contained in
all possible $p$-loop histories. If we recall that $0$-loop
corresponds to a point (0-D),$1$-loop corresponds to a line
($1$-D), etc. then  a total number of all the $p$-loop histories
in $D$-dimensions is $2^D$. As a result the respective Shannon
information $KLn(N)$ is equivalent to the number of dimensions
$D$ ( up to a non-essential constant factor). Information in turn
creates energy and is created by energy. This leads to a "triad"
of equivalence generated by a recursive loop: dimensions
$\leftrightarrow$ information $\leftrightarrow$ energy. The
recursive loop could be viewed as some kind of a spiral of
eternal creation and annihilation of energy, dimensions and by
implication geometry.

Another basic postulate of a new relativity is Nottale's scale
relativity \cite{ln1} takes the Planck scale $\Lambda =1.62
\times 10^{-35}$m as the ultimate ( and absolute) scale. This
postulate allows one to unify the spaces of different dimensions
similar to the unification of time and space in Einstein's
relativity ( on the basis of the absolute character of the speed
of light). Subsequently, a new relativity theory does not  need
to deal with compactification and decompactification of
superstring theory. The latter pose a serious problem since it
leads to a huge number of  possible phenomenological  theories of
the world. A new relativity theory adopting the
infinite-dimensional universe and the ultimate scale $\Lambda$
attainable in Nature suggests that topological field theories are
the most natural candidates for a "theory of the world".
Indirectly this is confirmed by the assumption that below Planck
scale there is no such thing as a distance ( interval) meaning
that the topology becomes a decisive factor there.

The third postulate of a new relativity is an extension of the
ordinary spacetime to noncommutative C-spaces thus making a
quantum mechanical loop equations fully covariant. This is done
by extending the concepts of ordinary spacetime vectors and
tensors to non-commutative Clifford manifolds where all p-branes
are unified with the help of  Clifford multivectors. There exists
a one-to-one correspondence between a nested hierarchy of
0-loop,1-loop,..., p-loop histories in $D$ dimensions coded in
terms of hypermatrices and single lines in Clifford manifolds.
This is roughly equivalent to Penrose twistor program. The
respective non-commutative geometry is taken to be the
transfinite continuum of Cantorian fractal spacetime studied
extensively by M.El Naschie \cite{en2,en1,en3} and G.Ord
\cite{go1}.

Recently a new relativity theory claimed a few successes. One of
those is especially interesting. A few years ago  \cite{ew1} the
leading figure in  string theory E.Witten remarked, probably with
some disappointment, that " a proper theoretical framework for
the extra term in the uncertainty" (stringy) "relation has not
emerged". It has turned that a full-blown stringy uncertainty
relation ( and its truncated version) could be straightforwardly
derived within the framework of a new relativity with the help of
its basic postulates \cite{cc3,cc5}. Also, in Appendix A we
provide an elementary heuristic derivation of the stringy
uncertainty relation based only on the postulate of the minimum
attainable scale ($ \Lambda$) in Nature.

In addition, it was shown \cite{cc4} that a new relativity
disposes of both  EPR and black hole information loss paradoxes.
More recently , it allowed one of us \cite{cc5} to derive in an
elementary fashion the black hole area-entropy relation valid for
any number of dimensions. In Appendix B we briefly review this
derivation and demonstrate that Bekenstein-Hawking relation is a
particular case of a more general relation. Furthermore, we argue
why the holographic principle is a direct result of a new
relativity.

A new relativity's most surprising prediction of group velocity
exceeding speed of light ( without resorting to tachyons) found
its  verification in recent experiments by Ranfagni and
collaborators \cite{Ran} who discovered the   evidence for "
faster than light " wave propagation to distances up to one meter.
\smallskip

In section {\bf 2} we consider an ensemble of dimensions
(including infinity) of quantum spacetime and derive its
distribution function. The subsequent calculation of the
respective average dimension associated with this distribution is
given in section {\bf 3}  where we obtain the value  for this
average which is very close to $4$ + $\phi^3$ ($\phi =
(\sqrt{5}-1)/2$ is the golden mean) previously obtained on the
basis of a Cantorian fractal spacetime model. A perceived
4-dimensional world then follows as a result of a coarse-grained
long range averaging effect of the underlying Cantorian fractal
geometry, as discussed by El Naschie \cite{en2,en1,en3}. In
addition we present a brief discussion of a plausible chiral
symmetry breaking mechanism in  Nature.
\smallskip

In section {\bf 4} we show why the "cosmological constant problem
"  is never an issue within the framework of a new relativity.
This follows from the idea of the automatic self-organization and
self-tuning of the universe  in accordance with the
renormalization group flow with respect to a local scaling
microscopic arrow of time \cite{ellis}( see Appendix C). We also
propose a cosmological scenario as an alternative to the big
bang, inflationary, brane-worlds and variable speed of light
cosmologies. According to our model, the world  began as a result
of a non-equilibrium process of self-organized critical phenomena
launched by vacuum fluctuations in Cantorian fractal spacetime.
Its future evolution proceeded in such a way that that determined
our existence in a metastable vacuum with a a subsequent
transition to the  renormalization group fixed point of a
dimension $ D$ = $4$ + $\phi^3$ \cite{en2}.

\smallskip
The above "fixed critical point" is not a true fixed point but
rather a metastable one, which means that it does not represent a
true vacuum. In section {\bf 5} we discuss how a new phase
transition will eventually drive the universe to another
(quasi-crystal) phase of a lower average  dimension of $<D>=
\phi^3$ representing the true vacuum. Using one of the basic
postulate of a new relativity ( Nottale's scale relativity) we
arrive at two integral expressions which determine the upper
limits imposed on a) the Hubble radius and b) the size of the
quasi-crystal phase. These expressions show that both limits
depend on the value $\phi$ of the golden mean which is a
recurrent feature of our analysis.

\smallskip
Finally, in the concluding section {\bf 6} we summarize our
results and write  down, for finite values of $D$, the unique
quantum master action functional for the world in {\bf C}-space,
(outside spacetime). This action functional governs the full
quantum dynamics for the creation of  spacetime, gravity, and all
the fundamental forces in nature. It appears that the quantum
symmetry in such a world should be represented by a braided Hopf
quantum Clifford algebra. The quantum field theory which  follows
from such an action is currently under investigation \cite{cc6}.
\bigskip

\bigskip
\section{Distributions :  Fermat's Last Theorem and a Multidimensional World }
\bigskip
To proceed with calculation of an average dimension of the
observable world we make an  assumption that there exists a
certain ensemble of universes with dimensions ranging from some
zero (reference) point( to be determined) to infinity. The above
reference point is analogous for example to a zero point energy
of a quantum oscillator. To simplify our analysis we choose a
spherical symmetry and represent the introduced ensemble as a set
of hyper-spheres whose radii are integer multiples of the
fundamental quantum of length, the Planck scale $\Lambda$. If we
would have chosen , say hyper-cubes instead of hyper-spheres then
it would have changed  the respective geometry of the ensemble
but not its topology. However the latter is more fundamental than
the former and therefore justifies a choice of the spheres as the
basic constituents of the ensemble.

 \smallskip
In what follows we pattern our derivation  after Planck's
approach to the black body radiation. The ensemble of the
universes is in contact with an information bath prepared by some
"universal observer" existing (we use the word $\it existing$ at
the wont of a more appropriate word) $\bf outside$ of the
ensemble. Such a frame of reference associated with the "
universal observer " can be associated with a hyperpoint of
infinite dimensions having an infinite amount of information (
see Appendix C) which could undergo a self-ordering (
bootstrapping) when perturbed ( even slightly).

 \smallskip
The members of the ensemble ( hyper-spheres, or p-branes) can
exchange energy ( and information) with the "walls" of this bath
at the Planck temperature $T=T_P= 1/\Lambda$ ( in units
c=$\hbar$=K =1) which is proportional to the number of
hyper-spheres, that is to the information content of the bath. We
assume ( in the spirit of things quantum) that the energy
exchange occurs in integer "chunks". Equivalently, the
hyper-spheres can absorb quanta ( from the bath) emit quanta ( to
the bath), and exchange quanta between themselves. These quanta
represent background independent geometric bits ( see Appendix B)
or the true quanta of spacetime and thus replace gravitons of the
linearized gravity which are not background independent. The
geometric bits form their own background where they continue to
evolve further. This is in agreement with one of the postulates
of a new relativity ( described earlier), namely the Chew
bootstrap idea. When the latter is applied to p-branes ( our
hyper-spheres) it states that all the p-branes are made of each
other.

 ince geometric bits are quanta  of fundamental excitations of
spacetime we do not need to have $\it a priori$ an embedding
target spacetime background where excitations of $p$-branes are
to propagate. The above ensemble is self-supporting (
self-referential),i.e. it obeys the bootstrap conditions. This
results in a background independent formulation of the emergence
of spacetime in a sense that the target background spacetime is
not fixed $\it a priori$. The ensemble itself creates (
self-reproduces) its own background in a process of the
evolution. Quite recently Smolin and Kauffman pointed that
self-organized critical phenomena \cite{ls1} might play a role in
the description of quantum gravity. In particular such processes
would have  allowed the emerging universe to self-tune its
fundamental constants, e.g., the cosmological constant.

\smallskip
Interestingly enough,there exists a connection between our
ensemble of universes and p-adic topology. In p-adic topology all
sets are simultaneously open and closed ( $\it clopen$), and
every point of the set is its center. This means that there is no
preferred point which could be considered as a reference point;
the only meaningful statements which can be made is about
relations between different points. If the introduced ensemble of
universes is also simultaneously opened and closed then there is
no preferred hyper-spheres in the ensemble: the only thing that
counts is their relations which nicely fits into Mach's principle.

\smallskip
Since there is no preferred "point" in the ensemble we are left
with only one choice for the reference frame, namely the thermal
bath prepared by the "universal observer" ( see above). Within this
frame of reference we can define the ground ( reference) energy, an
analog of the zero point energy of a quantum oscillator. To this end
we
assume that all the hyper-spheres in the ensemble  have the same
radius equal to the Planck length, $R =\Lambda$, that is the minimum
attainable length in nature. The respective Planck energy
$1/\Lambda$
could be taken as the reference energy.

\smallskip
The latter statement raises the following question. Do all the
hyper-spheres of the Planck radius but of different dimensionality
have the same value of reference energy? If we assume that the
respective energies are different we arrive at a contradiction,
since we have chosen the  reference point to be the  same for any
dimension. A choice of different values of the reference  energy
( according to the dimensionality) would violate the
poly-dimensional covariance (one of the postulates of a new
relativity) stating that all the dimensions must be treated on
the same footing ( there is no preferred dimension)\cite{wp1}.
Hence all the unit hyper-spheres in any dimension have the same
reference ( vacuum) energy  $\cal E$.

\smallskip
We postulate the invariance of the bulk energy density for
hyper-spheres of arbitrary radii $R_1,R_2,etc$ in any dimension
$D$, that is
 $$ { E_D (R_1)  \over C_D R_1 ^D } =    { E_D (R_2)
\over C_D R_2
^D } =... ={ {\cal E} \over C_D \Lambda^D }. \eqno (2.1)$$ where
$C_D= R^D{  \pi^{D/2}  / \Gamma ( {D+2 \over 2} ) }$, $E_D(R_j),
j=1,2,...$ is a radial excitation energy of a hyper-sphere of  a
radius $R_j$. The above postulate is based upon "incompressiblity"
(or  volume-preserving diffeomorphism symmetry) that appeared
within the context of $p$-branes on many occasions. The
hyper-spheres are quantized in integers of Planck units. An
increase in size of a $D$-dim hyper-sphere  in a process of
absorption of  energy "bits" occurs in such a way as to maintain
the energy density equal to the vacuum energy density, ${ {\cal
E} / C_D \Lambda^D }$

 \smallskip
 From eq.(2.1) follows that the energy $E_D$ is:

 $$E_D = { {\cal E} \over C_D \Lambda^D }C_DR^D =
 {\cal E} ( { R\over \Lambda} )^D.  \eqno (2.2)$$

 Thus we arrive at the following picture: the ground state of the
 initial ensemble comprises a condensate  of hyper-spheres ( at
 Planck temperature $T_P$) of all possible dimensions, where each
 hyper-sphere has a unit radius (in Planck units) and a constant
 energy ${\cal E}_{vac}$. A pioneering concept of a vacuum state
 of quantum gravity as a condensate was originally introduced by
 G.Chapline \cite{gc1}. A condensation at high temperatures was
 studied by Rojas {\it et al\/} \cite {r1}.

 \smallskip
 The condensation occurs in such a fashion
 that the  hyper-spheres in the ensemble tends to distribute
 themselves so as to  minimize  the energy  densities. For
 example, the spheres in the dimension range $D=4-6$ ( corresponding
 to the maximum volume of a unit sphere) will have
 a higher statistical weight than those in the extreme values
 $D=-2,~ \infty$ of the dimensions corresponding to a zero volume
 of a unit hypersphere.

 Using the principle of a minimum energy density we introduce
 the gamma distribution as the distribution function for the ensemble
 of hyper-spheres ( with the Planck radius normalized to one):

 $$ C_D = { \sqrt \pi^D \over \Gamma ( {D+2 \over 2} ) }. \eqno
 (2.3)$$ Here the values of $C_D$  are the statistical weights of
 the distribution. To account for thermal effects due to a
 decrease of the average temperature , from the Planck temperature
 to about 3 Kelvins we include into distribution (2.3) the energy
 dependence by using the Bose-Einstein distribution based on the
 hyper-sphere energy $E_D$ and its temperature $T$. Therefore the
 distribution density $\rho$ will take the following form $$
 \rho(D,E) = C_D /[e^{\frac{E_D}{KT}}-1] =C_D/[e^{\frac{{\cal E}
 (R/\Lambda)^D }{KT}}-1]\eqno(2.4)$$ where we use eq.(2.2).

 \smallskip
At the beginning of this section we postulated that the introduced
ensemble contains hyper-spheres of infinite dimensions. Now we
prove this statement ( at least for integer dimensions) with the
help of the Fermat's last theorem. Upon collision two
hyper-spheres, $A$ and $B$ can produce a third hyper-sphere $C$
analogous either to a final product of a chemical reaction or to
a {\bf 3}-point vertex in string field theory. Let us assume for
the moment that all the hyper-spheres have the same  dimension
$D$. The above interaction then conserves energy. Taking into
account additivity of the energy  we get

 $$E_A + E_B =E_C ~\Rightarrow (n_A)^D + (n_B)^D = (n_C)^D. \eqno
 (2.5)$$ where, after a cancellation of a common
 factor ${\cal E} $ we get the energies expressed as integers (
 quanta)
 $n_A, n_B, n_C$.

 \smallskip
 According to the Fermat's last theorem, eq.(2.5) has no  solution
 in nonnegative integers for integer dimensions greater than $2$.
 Since we live in the universe of the average dimension $D>2$ the
 Fermat last theorem requires an energy balance different from
 (2.5) : $$ (n_A)^{D_A}  + (n_B)^{D_B}  = (n_C)^{D_C}. \eqno
 (2.6)$$ where the dimensionalities $D_A, D_B, D_C$ cannot be all
 equal. Hence we have arrived  at one of the most important
 results of our work. According to the Fermat last theorem the
 equilibrium ( or quasi-equilibrium ) state of a  thermalization
 process  can be attained only if the dimensions of the colliding
 hyper-spheres must change in the process. Since dimension is a
 topological invariant, such a collision represents a simple
 example of a topology changing process while geometry is
 restricted to spherical geometry.\footnote{As an aside note we
 should mention that alternative geometries , for example
 hyperbolic geometries, such as the upper complex plane, de
 Sitter,and anti de Sitter spaces could be studied using a
 different framework. The problem with the respective topologies
 is that they comprised open and noncompact spaces. One could
 compactified them by attaching the projective boundaries at
 infinity. This would be topologically equivalent (in the anti de
 Sitter case) to spherical geometry $S^n$ thus reducing the
 problem to the previous case. We also would like to point out
 that a phase transition might not only change topologies but also
 transform one geometry into another.}

 \smallskip
 Let us consider an example of a simple dimension-changing process:
 $(2)^1 + (5)^2 = (3)^3$, that is $ A(n_A =2; D=1)  B (n_B =
 5;D=2) \rightarrow C(n_C =3; D=3).$ Here 2 quanta (geometric
 bits) plus  25 quanta ( geometric bits) yield 27 quanta (
 geometric bits). Since energy has been conserved in this process
 so does the information. The inverse process is also possible.
 State $C$ can decay into a sum $A+B$. This would look like some
 dimensional " compactification " process. A $D=3$ sphere of
 radius $3$ in Planck units has "compactified " into a sphere of
 $D=1$, of radius $2$ in Planck units,  and another sphere of
 $D=2$, of radius $5$ in Planck units. Using the analogous
 arguments we can consider more complicated collisions , like
 $A+B \rightarrow C+D$.

 \smallskip
 Since the dimensions' fluctuations are not necessarily
 small one has to abandon a perturbative expansion used in string
 theory.  For example, the perturbative string theory could not even
 reproduce a dimension change from $D=10$ to $D=11$, let alone
 handle infinity of dimensions. On the other hand, by using all the
 dimensions ( including infinite ones) on equal footing and
 employing the Fermat last theorem we bypass the need to use
 perturbative methods and an unpleasant task of summation over all
 the topologies. The  thermalization process of collisions
 (described by the Fermat theorem) automatically performs
 summation over different topologies. Thus, for example, $S^2$
 does not have the same topology as $S^4$ simply because the
 respective dimensions are different, the fact taken into account
 by eq.(2.6).

 \section{ Average Dimensions and Cantorian-Fractal Spacetime  }

 \bigskip

 \subsection{ Calculation of Average Dimensions }
 \bigskip

 Using the distribution density $\rho$ for the ensemble of
 hyper-spheres we get the average dimension of the ensemble as
 follows :
 $$ <D> = {\int^\infty _d dDdE ~D\rho (D,E)\over \int^\infty_d
 dDdE~\rho (D,E)  } . \eqno (3.1)$$ The lower limit of integration $d$
 must be found on the physical grounds. In a new relativity
 dimensions ( as everything else) are defined in relation to a
 reference point. From the dependence of either a volume or an
 area on dimensions follows that for any finite radius both the
 volume and the area tend to 0 at $D_0 =-2$. We take this value as
 the reference point. Another limiting value of $D$ resulting in
 zero volume and area is $D \rightarrow \infty$. These considerations
 determine the limits of integration in (3.1).

 \smallskip
 Since a hyper- sphere energy is given by (2.2) we immediately
 obtain
 $$dE = D {\cal E} ( {R\over \Lambda})^{D-1} d ({R\over \Lambda}).
 \eqno (3.2)$$
 where  $\Lambda$ (in units of $\hbar =c=1$) can be represented in
 terms of
 Newton's gravitational constant $G_D, G_{D-1}, G_{D-2},...$ in
 $D, D-1, D-2,...$ dimensions.

 $$\Lambda = G_D^{ {1 \over D-2}} = G_{D-1} ^{ { 1\over D-3}}
 =...\eqno (3.3)$$ In  $D=4 $ the Planck scale
 is the familiar $10^{-33} cms$. When $D=2$  the Einstein-Hilbert
 action is a
 topological invariant and (3.2) results in  singularity. We can
 avoid this by choosing $G_2=1$. This means that we can also choose
 the value of the universal  scale to be $1$ in the units of $\hbar =
 c =
 G_2=1 $.

 \smallskip
 Upon substitution of (2.4) in (3.1) we get

 $$ <D> = { \int^\infty_d dD~D^2  {\sqrt \pi}^D  [ \Gamma ({ D+2
 \over 2 }) ]^{-1}  \int^{(R_H/\Lambda)}_ 1  d(R/\Lambda)  ~
 (R/\Lambda)^{D-1}~ [e^{ {{\cal E}(R/\Lambda)^D\over KT}} -1 ]^{-1
 }                \over \int^\infty _d dD~D   {\sqrt \pi}^D [
 \Gamma ({ D+2 \over 2 })]^{-1}   \int^{(R_H/\Lambda)}_ 1
 d(R/\Lambda)~  (R/\Lambda)^{D-1} ~[ e^{ {{\cal E}
 (R/\Lambda)^D\over KT}}  -1 ]^{ -1 }} . \eqno (3.4)$$

 Evaluating the integral over$R/\Lambda$ we easily obtain :

 $$<D> = {    \int^\infty_d dD~D  {\sqrt \pi}^D [ \Gamma ({ D+2
 \over 2 })]^{-1}   F(a, b^D) \over \int^\infty _d dD~ {\sqrt
 \pi}^D  [ \Gamma ({ D+2 \over 2 }) ]^{-1}   F(a, b^D)} . \eqno
 (3.5)$$

 where $R_H$ is the Hubble radius whose value we {\it  do not
 fix\/} a priori and thus leave for the time being arbitrary.

 $$a = {{ \cal E} \over KT },~~~b = {R_H  \over \Lambda}, ~~~F(a,
 b^D) = a (b^D -1)  [ {  e^a  -1 \over e^{ab^D }   - 1 } ].
 \eqno (3.6) $$

 The average dimension given by (3.5) cannot be expressed in
 closed form. Therefore we will evaluate it numerically. Still we
 can extract some information directly from (3.5) before resorting
 to numerical integration. In fact, we can distinguish $3$ easily
 identifiable regions in terms of the parameters $a$ and $b$:

 (i) $a=1$. This means that  the thermal factor $F( a, b^D) \equiv
 0$ and thus does not influence the resulting value of $<D>$ leaving
 it as if it were no thermal factor.

 (ii)$a >> 1, b>> 1$. In this case the thermal factor $F( a, b^D)$
 once again tends to zero and will be factored out of the
 expression for $<D>$. The latter remains the same as without the
 thermal factor, with the only difference that the final
 temperature $T$ will be much lower than the Planck temperature
 $T_P$ corresponding to the vacuum energy $\cal E$.

 (iii)  $a > 1$, $b>1$. In this case we cannot factor out the
 thermal factor $F( a, b^D)$, and as a result it contributes
 significantly to the value of $<D>$. Numerical simulations show
 that in this case the fluctuations in $R_H/\Lambda$ and $T$ will
 initially lead to an increase of $<D>$ and then ,after reaching a
 maximum value, to its monotonic decrease approaching
 asymptotically the metastable fixed point $<D> =4+\phi^3 =
 4.236...$ of the renormalization group. This point corresponds to
 our world which emerged from chaos ( a multitude of dimensions)
 as soon as the ensemble of the hyper-spheres began evolving
 (generating the respective average temperature and dimensions)
 towards the metastable fixed point of the renormalization group

 \smallskip
 As a next step we will provide the numerical value of $<D>$ and
 demonstrate an importance ( and ubiquitousness) of  the golden
 mean $\phi$ in evaluations of the average dimensions of the world.
 In particular, it will help us to understand why we live in four
 dimensions, why their signature is $3+1$, and why there might
 exist a deep reason of chiral symmetry breaking in nature.

 \bigskip

 \subsection{The Cantorian number $<D>~ =4 + \phi^3 = 4.236067977...$
 as
 the Exact Average Limiting Dimension }
 \bigskip

 We have already mentioned that a new relativity requires that
 dimensions should be calculated with respect to the reference
 point ( zero point dimension)$D_0$ , that is any dimension $D$ is
 in fact $D' = D-D_0$. In the previous section we have already
 chosen this reference point $D_0 = -2$ Introducing $<D'>$ into
 eq.(3.5) we get

 $$<D'(t) >  ~=  {  <D' F(a (t) , b(t)^{ D'} ) > \over < F(a(t) ,
 b(t) ^{ D'} )> } ~ \ge~ <D'>_{lower} . \eqno (3.7)$$

 Here  $<D'>_{lower}$ is the average dimension without quantum
 dissipative effects (responsible for the initial chaotic phase).
 The latter will increase the average dimension as compared to
 $<D'>_{lower}$ and prevent the expanding hyper-spheres  from
 cooling down below the Planck temperature $T_P$. As soon as the
 average dimension reaches a peak value, the continuing expansion
 of the hyper-spheres will gain the "upper hand" over  the
 re-heating due to quantum-dissipative effects. As a result the
 hyper-spheres will begin to cool, and the average dimension will
 begin to decrease towards its initial value. This will signal an
 onset of the  " ordered " phase emerging from the initial chaotic
 phase (quantum dissipation ). It is possible to demonstrate that
 the average dimension cannot fall below the metastable fixed point
 $4+\phi^3$ ( metastable vacuum) of renormalization group until the
 system will reach the phase transition point and move towards the
 quasi-crystal phase ( the true vacuum ).

 \smallskip
 Evaluating expression (3.5) for cases (i) and (ii)we easily find
 that
 $$<D'> ~= 6.3876 =~ <D-D_o> ~~~ \Rightarrow
 <D> ~= 4.3876 \sim
 4+ \phi^2 = 4.3819.....\eqno (3.8)$$

 This value represents the upper value of $<D>$. The respective lower
 value can be evaluated as follows. Assuming that the dimensions
 are integers we replace in (3.5) the integrals by the sums and
 obtain
 $$<D_n> = 4.3267 \eqno(3.9)$$
 which is smaller ( as expected ) than the value given by (3.8) ,
 that is $$4.3876 \sim 4 + \phi^2 = 4.3819 ... \eqno(3.10)$$

 The values of $<D>$ based on a discrete (3.9) and integral(3.8)
 averages are in a very good agreement with El Naschie results obtained
 with the help of  the transfinite $\cal E_\infty$Cantorian fractal
 spacetime models in \cite{en3},\cite{en5}. These models give the
 following exact value of the average dimension :

 $$<D> = < dim_F ~ {\cal E} ^{ ( \infty) } > =  { 1 +\phi \over 1
 - \phi } = {1 \over \phi^3} =  4+ \phi^3 = 4.236... \eqno (3.11)$$
 where $\phi = { \sqrt 5 - 1 \over 2}= 0.618... < {ln 2/ ln 3}$
 and $ ln 2 / ln 3$ is the dimension of the middle segment Cantor
 set .

 \smallskip
 The golden mean $\phi$ provides us with the basic "unit", or the
 elementary dimension building block from which all the sets of the
 transfinite Cantorian fractal spacetime models are constructed.
 These spaces are the densest spaces obtained from infinite
 intersections of infinite unions. The value of $ d_c^{(0)} = \phi=
 0.618033984....$ is the fractal dimension of a physical structure
 living in a $0$ topological dimension. This can be interpreted as
 packing, compressing ( information, for example  ) the fractal
 dust points of dimensionality equal to $\phi$ into a " point ".
 We put quotation marks around  " point " to emphasize that since
 there exists the minimum attainable (non-zero)  scale in nature (
 Planck scale) there are no  points in nature. A geometrical point
 is in fact smeared out into a " fuzzy " ball/sets  of all possible
 topological dimensions, from $-\infty $  to $ \infty$. For this
 reason the naive concept of a  fixed and well defined dimension
 cannot be used anymore.

 \smallskip
 The  set of $-\infty $ dimensions ${\cal E}^{(-\infty)}$  is
 called the void virtual set and the set of $\infty $ dimensions
 ${\cal E}^{ (\infty)}$ is called the universal set.  The void set
 has in fact the fractal dimension $0$, i.e the void set comprises
 one single fractal dust point. The universal set,on the contrary,
 has an uncountable infinity of fractal dust points which means
 that its fractal Cantorian dimension is $\infty$. Thus the
 respective dimensions are dual to each other : $d_c^{ ( \infty)}
 = \infty$, $d_c^{(-\infty)} = 0= 1/d_c$.

 \smallskip
 The virtual sets of negative topological dimensions can be
 endowed with a physical meaning if we would view them as carriers
 of negative information entropy, or anti-entropy \cite{en4}. Such
 an interpretation is very close to Dirac's sea of negative energy
 where we substitute negative entropy for the negative energy. The
 true vacuum of the theory requires the sea of negative
 topological dimensions located below the topological value of a
 zero dimension $D_0 =-2$ to be filled up. This is necessary for
 keeping the quantum universe from "cascading" down and
 disappearing in a void set $\cal E^{-\infty}$.

 \smallskip
 Thus we arrive at the following result: the average dimensions
 obtained with the help of  the Cantorian fractal spacetime model,
 discrete sums average,  and the integral average respectively
 satisfy the following relation :

 $$  4+\phi^3  = 4.236... ~ <  4.3267~(discrete )   ~< 4.3876 ~
 (integral) \sim 4 + \phi^2.   \eqno (3.12) $$ The difference
 between various averages is surprisingly small, taking into
 account that our model uses smooth spheres  and that $4$+$\phi^3$
 is the exact result obtained by El Nashie on the basis of the
 densest packing allowed and which generalizes the
 Maudlin-Williams golden mean theorem. The spaces are filled-in
 totally with " fractal dust points ". Another values for the
 average dimension $<D > = - 2 / ln \phi = 4.156...$  (based on
 the knot theory and Jones polynomial \cite{ln1},\cite{en3})  and
 $<D> =\sim 4.07 -4.09$ (based on the Leech lattice packing) are
 only an approximation to the exact value of $<D>$=$4$+$\phi^3$. This
 is due to the fact that these models used packing different form
 the densest packing.

 \smallskip
 We believe that connection of the exact average dimension of the
 world to the golden mean is not a simple numerical coincidence.
 In fact, this value  $<dim ~ {\cal E}^{\infty }>= d_c^{(4)}$ =
 $4$+$\phi^3$  indicates the onset of quasi-ergodicity
 \cite{ln1},\cite{en3}, that is the topological dimension is of
 the same order as the Fractal dimension . $D_T > D_F$ represents
 a stable region and $D_T < D_F$ represent ergodic case. We are
 living in a metastable quasi-ergodic state or " vacuum ", $<D>$
 =$4$+$\phi^3$. On the other hand, the true vacuum, as was argued
 earlier (in connection with $\cal E^{(-\infty)}$ and $ \cal
 E^{\infty}$) corresponds to the dual inverse fractal dimension of
 $d_c^{ (4)}$ equal to $d_c ^{ ( -2) } = 1/ ( 4 + \phi^3)=
 \phi^3$; or $ 2 + \phi^3$ relative to the zero point $ D_0 = -2$.

 \smallskip

 Still the major question remains unanswered: why do we live in a
 $3 + 1$ dimensions? To answer we have to consider the long
 range coarse-grain averaging process of a  fractal geometry
 underlying the perceived spacetime of our existence.

 \smallskip
 Such an averaging process results in a breakdown of
 poly-dimensional isotropy. This means that prior to the break-down
 there are (on average) four orthogonal dimensions which
 separately can deviate from their respective values by the same
 amount $(\epsilon/4) = (\phi^3/4)$ . As the result the emerging
 average dimension is

 $$4 ( 1 + {\phi^3 \over 4} ) =  4 +\phi^3 = 4.236...=
 <dim_F~ {\cal E}^{ (\infty)}> = dim~{\cal
 E}^{(4)}= d_c^{ (4)}. \eqno (3.13)$$

 After the breakdown of isotropy  $3+\epsilon = 3(1+\epsilon/3)$
 spatial dimensions plus $ 1+ \epsilon $ temporal dimensions
 emerge. Here the fluctuations of the temporal dimension is 3
 times greater then the fluctuation $\epsilon/3$ of each of the
 spatial dimensions.The long-range coarse-grain averaging
 process could be viewed as a process of "projecting " El
Naschie's effective average $4$+$\phi^3$-dimensions onto the
four-dimensional
 $\it smooth$ outer surface, which we perceive as our reality. As
 a result of the projection the initially orthogonal spatial and
 temporal dimensions will become entangled. A good analogy could be
 a two-dimensional view of a three dimensional knot.

 \smallskip
 This could explain a perception of the three-spatial dimensions,
 at any given fixed moment of clock time. One can slice ( at least
 locally) the four-dimensional spacetime into  $3$-dimensional
 sheets at any given fixed value of the clock time resulting in a
 possibility of  measurements in extended space at a fixed moment
 of time \cite{dc1}. Inversely, it is difficult ( if not
 impossible)  for a macroscopic observer to perceive an extended
 interval of time, within an extended region of  $3$-space.
 Paraphrasing Wheeler \footnote {"Time exists to prevent all things
 from happening at once"}, we can say, " space exists to prevent
 everything from happening one point ".

 \smallskip
 On the other hand, in our model of infinite dimensions at the
 scales approaching Planck scale particles can move forward and
 backward in time simply because they have access to more
 dimensions than we do. In other words, particles are fully "aware" of
 the fractality of spacetime. For example, they do not "see" the
 two slits in the two-slit experiment as two separate  points
 since from their perspective spacetime is a discrete, fractal-like
 Cantor set. When one constructs the Cantor set by removing the
 middle segment $ad~ infinitum$ , one is literally " removing "
 the space between the fractal dust points thus eliminating a
 separation  between the slits. Since at this scale there is no
 separation between the fractal dust points, the Cantor set has
 zero measure = zero length. It is our mind which  fills the void
 creating the illusion that there is a separation between the
 points of a fractal dust. This leads to the apparent "
 non-locality " and other paradoxes of quantum mechanics. Nature
 does not abhor the vacuum , it is our mind which does so by
 filling in the voids. In fact, nature is not only quantum and
 discrete, it is fractal.

 \smallskip
 Thus the entanglement and long range averaging of the perturbed
 spatial and temporal dimensions yield the observed $4$-dimensional
 world. This means that within the framework of Cantorian fractal
 spacetime  we have the following entangled dimensions yielding
 the perceived $4$ dimensions

 $$ (3+ \epsilon ) ( 1+ \epsilon ) = 4 ~\Rightarrow 4+\epsilon =
 {1\over \epsilon } ~ \Rightarrow  \epsilon = \phi^3 . \eqno (3.14
 )$$ As a result of this process the golden mean appears once
 again as an essential parameter of reality. On a surface this
 looks like a forced fit to the preconceived notion. And really,
 why does not $\epsilon=\phi^3$ appear as a result of some other
 splitting, say $(5+\epsilon_1) (1+\epsilon_2) =6 $? . It is
 obvious that there are infinite number of solutions $\epsilon_1,
 \epsilon_2$ satisfying suac splitting. However, to require that
 dimensional perturbation  $\epsilon_1=\epsilon_2= \phi^3$ must be
 the only solution indicating the emergence of $3$-dimensional
 entangled imposes a very stringent constraint. We must mention
 that the condition $\epsilon_1=\epsilon_2= \phi^3$ satisfies the
 symmetric split $(2+\epsilon) (2+\epsilon)=5$ between $2$
 spatial, $2$ temporal, and the universal fluctuation $\epsilon$.
 Is all this just a numerical coincidence or design ?

 \smallskip
 We can explain this situation from the lattice point of view. If
 we consider a quasi-ergodic self-similar fractal grid/lattice, or
 topologically equivalent  fractal sphere  of dimensionality
 $4$+$\phi^3$  the coarse-grain averaging process will break down
 the "translational " symmetry from

 $$ {\cal T}[4+\phi^3 ] \rightarrow {\cal T} [4] = T[(3+\epsilon
 )(1+\epsilon)]~~ where ~~~\epsilon = \phi^3. \eqno (3.15)$$

 It transpires that our perception ( which by the very definition
 amounts to averaging)  of $3$ dimensions is in fact the long range
 coarse-grain effect of co-existence of $3+\phi^3$ (spatial )  and
 $1+\phi^3$ (temporal ) dimensions. The set ${\cal E}^{(4)} $
 whose dimension is $4$+$\phi^3$  is "packed" inside a
 four-dimensional sphere. This reminds a picture of the grains of
 sand on a beach. The sand "looks"  two-dimensional to us when in
 fact it is $3$-dimensional. Due to a coarse-graining effect we as
 macroscopic observers can only perceive the "skin ", or a
 $3$-dimensional  outer surface (where space and time co-exist) of
 the "onion"-like world. In other words, we perceive only
 projections onto the " four-dimensional external surface " of the
 average $4$+$\phi^3$ dimensions of a truly infinite-dimensional
 world lying figuratively speaking  underneath our "feet" and
 above "our heads". The four-dimensional "skin"  encloses the
 exact  average value  of $<D>$=$4$+$\phi^3$ where averaging is
 performed over all possible topological dimensions (ranging from
 $-\infty$ to $\infty$) of the sets
 ${\cal E}^{ (i)} $ of Cantorian fractal spacetime. A similar mechanism may be
 responsible for the chiral-symmetry breakdown in Nature without
 supersymmetry.

 \smallskip
 If there will be enough energy to move "inside" the "skin"  into
 the transfinite continuum of Cantorian spacetime we will be able
 to discover the true fractal average dimension of spacetime, at
 each layer(ladder)  of the renormalization group flow. From the
 renormalization group treatment of quantum field theory one can
 infer that the universe, upon reaching the metastable fixed point
 $4+\phi^3$ of the renormalization group  will signal the
 beginning of a dimensional phase transition, from $4$+$\phi^3 $ to
 $\phi^3$. At this fixed point a conformal invariance exists which
 means that the world  will become self-similar at every
 scale.

 \smallskip
 From this perspective it seems that theories of gravity and all
 other fundamental forces in nature are effective theories since
 they  emerge from a deeper theory , namely a new relativity in
 Cantorian fractal spacetime. In fact, vacuum
 fluctuations in Cantorian fractal spacetime generate the long
 range Einstein's gravity theory and all other fundamental forces
 "residing" inside spacetime. It has turned out that a new
 relativity allows one to write down the quantum master action
 \cite{cc1,cc3} for  the "master" field residing in {\bf C}-space,
 ( outside spacetime) whose vacuum fluctuations generate classical
 spacetime, gravity and all  other fundamental forces. We believe,
 this is a reason why attempts to quantize Einstein's gravity have
 been futile so far. It looks as if Cantorian fractal geometry is
 the bridge connecting both worlds, quantum and classical.
 Therefore noncommutative geometry together with the
 renormalization group approach \cite{dk1} are  mathematical tools
 allowing us to probe this Cantorian fractal world \cite{en5}.

 \smallskip
 Thus if general relativity required four-dimensional Riemannian
 geometry for its formulation a new relativity requires the
 transfinite continuum of the noncommutative von Neumann's
 Cantorian fractal spacetime ${\cal E}^{ ( \infty)}$
 \cite{en1}-\cite{en3}. What we perceive as a smooth
 four-dimensional topological space ( a sphere, for example ) is an
 illusion. There are no " points " in this New Relativity, due to
 the fact that the Planck scale is the minimum distance in Nature.
 As we zoom in using a "microscope" of the renormalization group, we
 uncover that each  point is also a four-sphere, and that each
 point within that sphere is also four-dimensional, and so on and
 so forth...$ad~infinitum $.

 \section{A Solution to the Cosmological Constant
 Problem }

 \bigskip

 Here we will present a solution to the cosmological " constant"
 problem which parallels Nottale's derivation based on his scale
 relativity as well as El Naschie's $\cal E^{\infty}$ Cantorian
 spacetime theory. Depending either on the values of the
 "constants" $a,b$ of the previous section or initial conditions, one will have
 several different cosmological scenarios. We write "constants"
 because in reality they have an explicit scaling " temporal "
 dependence consistent with the renormalization group. The
 parameters $a, b$ will change with  scaling  time, $t$. We set
 the initial scaling  time to be $t_0 =1$ , since $ln (t_0)=0$ and
 logarithmic dependencies play an important part in scale
 relativity.

 Using (3.6) and the scale relativity we get

 $$a(t) = ({ {E_0}\over KT}) (t) = ( { {\cal E} \over KT_P }
 )^{(t/t_o)^\alpha }.~~~~ \Rightarrow { log~[ E_o /K] \over log~[
 {\cal E} /KT_P ] } = (t/t_o)^\alpha. $$

 $$b (t) =  ({R\over \Lambda}) (t) =( { {\cal R} \over \Lambda
 }   ) ^{ (t/t_o)^\beta }. ~~~~ \Rightarrow { log~[ R/\Lambda ]
 \over log~[ {\cal R} / \Lambda  ] }  = (t/t_o)^\beta   \eqno
 (4.1))$$
Since we introduce the time dependence of the parameters
 $a$ and $b$ we have two different values of the vacuum energy:
 $E_0$ denotes the vacuum energy at a given time $t$ and $\cal E$
 denotes the initial vacuum energy. The average value of the
 dimension is respectively:

 $$ { <D (t)>     \over  <D (t_o) >       }  = ( {t\over t_o}
 )^{-\delta}.~~~\Rightarrow   {  \partial ln~<D(t)> \over \partial
 ln~t        } = -\delta. \eqno (4.2)$$

 where $\alpha$ and $\beta$ are the constants entering
 Nottale's scale relativity \cite {ln1}, and $\delta$ is the
 universal scaling exponent. Equation (4.2) mean that
 the average dimension changes monotonically with scaling time. On
 the other hand, eq.(4.1) postulates a power-scaling law for the
 ratios of the fundamental constants in nature. According to
 Nottale's scale relativity the scaling coefficient themselves are
 resolution-dependent. However for the simplicity sake we assume
 that they are universal constants.

 \smallskip
 In Appendix B it is shown that the slope $dS/dA$ of entropy vs. area
 diverges between the values $D = 4 $ and $D =5$ . This represents an
 onset of a dimensional phase transition, that is the universe has
 reached $ <D> = 4+\phi^3 $ and begins its transition to the
 quasi-crystal phase characterized by $ <D > = \phi^3$ (or to
 $2+ \phi^3$ if we take  $ - 2$ as zero reference point). To study
 this process in more details one needs to consider Nottale's scale
 relativistic resolution-dependent coefficients $\alpha$ and $\beta
 $. At the critical point  these three coefficients are
 connected by a certain relation which is typical for critical
 phenomena.

 \smallskip
 For example if initially, at the Planck era the vacuum energy
 ${\cal E}$ was of the same order of magnitude as the thermal
 Planck energy $KT_P$ then generally speaking its value will
 subsequently change  (with the renormalization group flow) to a
 much smaller value $E_o$ of today. However ${\cal E}$ still will
 be large in comparison with today's  background thermal energy of
 $2.76-3$ Kelvins. We would like to reiterate that in a new relativity
 it makes sense to speak only about relations between quantities (
 and by implication about the respective ratios). In this sense a
 concept of a very small is intrinsically connected with a concept of a very large.

 \smallskip
 The quantities  ${\cal E}, {\cal R}$ are respectively the vacuum energies and
 sizes of the original baby universes at the " initial"  scaling
 time of $t_0 =1$, at the beginning of the "count down". This
 means that the scaling "clock " starts ticking at the moment when
the radius of the hyper-sphere is $ {\cal R } $ and the value of
the vacuum energy is ${\cal E} $. After that moment infinitesimal
perturbations of both, the radius and energy  will begin to grow.
This results in the following physical picture.

Initially the vacuum fluctuations about the perfect balance
conditions $(R/\Lambda)$ = $E/{KT_P}$= $1$ were infinitesimally
small. However, because the system has an infinite degrees of
freedom these fluctuations could not  have been damped, and this
would have lead to a disruption of equilibrium. Inversely, had
the number of degrees of freedom be finite the infinitesimal
perturbations would have been damped and equilibrium would have
been restored.

Thus, as a result of infinite number of degrees of freedom the
universe was driven out of balance, out of the metastable state
of average dimension $4$+$\phi^3$, to states  of higher and higher
average dimensions, until it reached a maximum  average dimension
which depends on initial perturbations and the non-linear dynamics
of the system.After reaching the maximum average dimension , the
universe began its "descent" from this maximum to the initial
value of $ D$ = $4$ + $\phi^3$.

Since  the initial perturbations drove the average dimension to a
value higher than $4$ + $\phi^3$ one might be inclined to think
that the average temperature will  also surpass the Planck
temperature $T_P$.However this is a wrong conclusion since
according to a new relativity the effective values of the
Boltzamann constant $K_{eff} (E)$ and Planck constant
$\hbar_{eff}(E)$  are  energy dependent ( see e.g, Appendix A).
Both constants increase in such a way as to keep the temperature
below $T_P$. For example, a very energetic photon of frequency
$\omega$ being emitted by the walls of the reservoir cannot
exceed the upper temperature bound of $(\hbar_{eff} \omega /K
_{eff} )  = T_P$.

 \smallskip
 As we discussed earlier, during the initial chaotic phase the
 quantum-dissipative effects re-heat the expanding universe in
 such a way as to keep $T_P$ constant. After the universe had
 reached its maximum average dimension, it began to "roll  down",
 in both dimensions and temperature: the ordered-phase began. The
 renormalization group local scaling arrow of time also appeared at
 the moment when the universe started its descent back to the
 metastable point. Prior to that there was no arrow of time since
 when the universe was in perfect balance nothing changed,
 nothing  "happened ". Spacetime as we know it did not exist and
 emerged only  afterwards. The universe began due to a
 non-equilibrium process of self-organization as was argued long
 ago by Prigogine.

 As the Universe moves from higher average dimensions and higher
 vacuum energies , to lower dimensions and lower vacuum energies,
 information is lost. Since the  total  entropy of the universe
 plus reservoir cannot decrease, this means that entropy flowed
 out of the universe into the reservoir which  comprises  the
 infinite family of hyper-spheres of all possible dimensions and
 radii. The quanta of spacetime create their  own background in
 which all the $p$-branes live in ; this background is
 self-referential and self-supporting.

 Keeping in line with the second law of thermodynamics, the
 information entropy  which flows out of a  universe into the
 self-referential thermal background is  recycled over and over
 again  due to the bootstrap principle , i.e., the
 quanta/hyper-spheres
 are made of each other. This recycling process of
 information-energy-dimensions bootstrapping, the creation of the
 fundamental particles in the universe and its life forms occurs
 in a hierarchical multi-fractal fashion. Accordingly
 self-organization proceeds in discrete jumps from a smaller scale
 to a larger scale, to a larger scale, etc. Therefore we subscribe
 to  Penrose's view that consciousness is a non-algorithmic process
 which is compatible with our idea that an uncountable infinity of
 dimensions implies existence of an uncountable infinity of
 information leading to the emergence of conscious life.

 \smallskip
 As the emerging local scaling time begins its "count" the values
 of the "universal constants " begin change with "time".
 This process  is consistent with a modern theory of variable
 speed of light cosmologies which gains a certain support. Choosing
 a simple situation characterized by  $ \alpha = \beta $ and ${
 {\cal E} / K T_P } = { {\cal R} / \Lambda }$  we get from (5.1)

 $$ {  ({ E_o/ KT}) \over   ({R/ \Lambda}) } (t) = [ {  {  {\cal
 E} /  KT_P } \over { {\cal R} /  \Lambda  }   }   ]
 ^{(t/t_o)^\alpha } = 1. \eqno (4.3)  $$  From Eq.(4.3) follows that
 in this simple case  for all values of renormalization group
 scaling time the following is true

 $$     ({R\over \Lambda})(t)  =  { E_o \over K T} (t). \eqno
 (4.4)$$

 Taking, for example, $R_{Hubble} = 10^{60} \Lambda $ we find from
 (4.4) that according to this simple scenario the present-day
 value of the vacuum energy should be

 $$  { E_o } (today) =  (kT_{H}  )~ ({R_H\over \Lambda})(today)  =
 10^{60} (kT_{H}).  \eqno (4.5)$$ This result means that the
 present vacuum energy would be  huge  as compared  to the thermal
 energy  $T_H$. This might serve as a very straightforward
 explanation of the fact that  today's universe  is expanding much
 more rapidly than was predicted by the existing theories. The
 explanation follows as a  natural result of a new relativity
 theory and Cantorian fractal spacetime which in turn might
 be viewed as an indirect proof that the respective world view  may
 be correct, or at least points in the right direction.

 \smallskip
 Let us look at the famous experiments $COBE$. The $T_{COBE}$
 value of $2.76-3$  Kelvins was obtained by "looking" into the past
 of the universe. The photons we detect today were emitted in the
 past and were redshifted due to the expansion of the universe.
 Since the speed of light in a new relativity changes with the
 renormalization group flow, calculation of  the exact values of
 the redshifts  is going to be a rather difficult task. The
 correct way to do this would be to carry out integration along
 the renormalization group trajectory backwards in scaling time
 (cf. use of convective derivatives in fluid mechanics) to find
 out  the true  frequency of the  photons upon emission.

 \smallskip
 In the widely accepted theories "past" refers to  the Big Bang.
 On the other hand, there was no Big Bang in our cosmological
 scenario. Our model implies an ever expanding Universe where
 expansion begins from the moment when the infinitesimal perturbations
 drove the Universe out of balance. By expansion we mean a
 synchronous change of universe's radius with the renormalization
 group flow representing the true arrow of time.\footnote{Here we must
 add a word of caution concerning the use of
 the true " local scaling " renormalization group flow arrow of
 time. It should not be confused with the conventional coordinate
 time. H. Kitada \cite{ht1}using Godel's incompleteness theorem has
 proven that a local  time exists even in the absence of global
 time in traditional quantum cosmology.}

 Since the average dimension of the universe changes in the process it
 leads to a
 change of  the respective volume even for a fixed radius. However
 to take the volume of the hyper-spheres as an indicator of
 temporal evolution is not a good idea. The radius is the more
 appropriate indicator of change. Quite analogously to the radius being an indicator of the
 renormalization group flow, the COBE data's temperature $T_{COBE}
 $ may also serve as an appropriate " thermometer " of the
 evolving ensemble of an infinite number of bubbles/universes. For
 this reason, in this model, one could set $T_H = T_{COBE}$ and
 conclude that within the model the vacuum energy today would be
 of the order $10^{60} KT_{COBE} $. Another way to look at the COBE
 data is to look for  deviations from the $2.76 $ Kelvin  due to a
 greater redshifts caused by a greater expansion of a cooling universe.
 Recent experimental work by De Bernardis {\it et al\/} and the recent
 BOOMERANG data indicates that this is the case \cite {b1, b2}.

 \smallskip
 The renormalization group scaling time changes rather  slowly as
 compared to the coordinate time of our clocks since the former is
 logarithmic  and the latter is roughly linear. In turn the
 fundamental constants of nature change with respect to the
 renormalization group scaling time, and for this reason they
 change very slowly as compared to our daily experience. For
 example, the speed of light at the time of relativity formulation
 was essentially the same as today. However,
 this was not the case when the universe began its evolution at the
 Planck era. At that period the fundamental constants changed more
 rapidly. We are living now in a metastable phase( a slow,
 "predictable") and for this reason life was possible at this stage
 of evolution.

 \smallskip
 It would be hard( if not impossible) for life forms
 to emerge during a phase where the fundamental constants changed
 rapidly relative to the renormalization group scaling time flow.
 Such a rate of change corresponds to the situation where the
 information-entropy is no longer a linear function of the area
 which in turn is a result of taking into account the
 infinitely-dimensional picture of the universe ( Appendix B). As
 for ourselves, we are living in a linear world where
 information-entropy transfer per unit area is constant. This is
 another reason why $ D = 4, 5  $ are the optimum dimensions
 conducive for the appearance of humans, which is consistent with
 the average dimension $D = 4$.

 \smallskip
 Let us look at the cosmological constant $\lambda$ in today's
 $4$-dimensional world. It has units of a $(length)^{-2}$ because
 it must have the dimensions of curvature appearing in
 Einstein's equations. The units of $\lambda $ are also the same as
 the units of a string tension, namely energy per unit length. Let
 us now compare today's $\lambda$ with the cosmological "constant" at
 the moment when the scaling renormalization group time was
 launched (at the Planck era) by simply evaluating the ratios of
 energy/length. One can argue that at the very beginning  the large
 scale structure of spacetime ( as we know it ) did not even
 exist. Spacetime was barely born at that moment . Therefore
 Einstein's equations could not be applied and this period is a
 truly quantum gravity period. Still we can compute the relevant
 numbers, at least crudely.

 Setting the vacuum energies to be of the same orders of magnitude
 as their  Compton energies , $ E_{today } \sim { 1\over R_H } $
 and $ {\cal E } \sim {1\over \Lambda } $ and using eqs.(4.3-4.5)
 we  arrive at the following ratios of the cosmological "constants" :
 $$  {  { E_{min} /  {\cal L }  } \over { { \cal E} / {\Lambda
 }        } } = ( { {\Lambda } \over {\cal L} } )^2 =  { T_o \over
 T_P }.\eqno (4.6)$$

 Let us assume for the sake of the argument that one could take
 the upper Nottale scale to be $\cal L $ $\sim  10^{61}\Lambda$ $>$
 $10^{60}\Lambda $. Using this assumption in (4.6) we obtain the
 following ratio of the cosmological " constants " :

 $$   ( { {\Lambda } \over {\cal L} } )^2 =  10 ^{ -122 } = { T_o
 \over T_P }. \eqno (4.7) $$ This result implies that the
 cosmological " constant " (when the world was of the Planck
 radius - a mini-black hole ) was  122 orders of magnitude larger
 than the cosmological " constant " of the universe having the
 maximum radius of ${\cal L } $. Once again we wish to emphasize
 that we are referring to the radius of the universe since in
 general dimensions change and therefore it is wrong to compare
 volumes at different epochs.

 \smallskip
 At another limit the universe would move to an extremely cold
 world of minimum $T_o$ which is close to zero but  not zero. In
 fact if the minimum $T=0$ this would contradict the duality of
 nature requiring that maximum $T_P$ must be  dual to a minimum $0$
 temperature since in this case $1/0 = \infty$. The state
 corresponding to $T_o$ is the noncommutative quasi-crystal phase
 of the coldest world whose dimension $d_c^{(-2)} = \phi^3 =
 0.236..$. This means that as the universe reaches the
 renormalization group metastable point of $ D = 4.236...$ it would
 begin to move very slowly towards another phase transition and
 not less slowly proceed to a very cold world whose lowest
 temperature $T_o$ will be 122 orders of magnitude smaller than
 the Planck temperature $T_P$ that is  $T_o \sim 10^{ 32 }. 10 ^{
 -122} = 10^{-90}$ Kelvin.

 \smallskip
 At a first glance it seems that the assumption about inverse
 proportionality of vacuum energy and the scale (that is $ E \sim
 1/R $) contradicts eq.(4.4) which states that the energy is
 proportional to the scale.  However after performing the duality
 transformation $T_o \rightarrow T_P,~~~{\cal L} \leftrightarrow
 {\Lambda }$ in eq.(4.4) and leaving ${\cal E}_{min}$ unchanged we
 obtain

 $$ { {\cal E}_{min} \over KT_P  } =  { \Lambda  \over
 {\cal L} } . \eqno (4.8) $$  It is immediately  seen that eq.(4.8) is
 in full agreement with the statement that ${\cal E}_{min } \sim
 1/ {\cal L} $. Moreover, combining eq.(4.4) and (4.8) we arrive at
 eq.(4.7). Interestingly enough, if we apply the duality
 transformation to a bubble of Planck radius, and Planck
 temperature $T_P$, we find that the Planck scale is " self dual "
 in the sense that the Compton wavelength associated with the
 Planck's mass $M_P$ is in excellent agreement with the value of
 its Schwarzchild radius , $R \sim M_P$. But this is precisely how
 one defines the Planck scale from the very beginning.

 \smallskip
 As we have already indicated, the universe will keep expanding
 until it will asymptotically approach the upper Nottale scale
 $\cal L$, reaching the final quasi-crystal phase. The scale
 relativistic corrections prohibit scales exceeding $\cal L$. This
 restriction serves as the natural infrared regulator which is
 analogous to viewing the Planck scale as the natural ultraviolet
 regulator. The relation between the two represents the duality  of
 UV/IR. On the other hand, the ultimate scale $\cal L$ corresponds to the
 lowest temperature $T_o$ dual  to the maximum attainable
 temperature $T_P$ ( a postulate of thermal relativity). This means
 that the universe's evolution  from the minimum Planck  scale $
 L$ to the Nottale's maximum scale ${\cal L} $ is accompanied by a
 change of the temperature from the maximum temperature $T_p $ to
 the minimum (dual) temperature $T_o $.

 \smallskip
 Between these 2 limits the universe can "hover" in the metastable
 state ( $<D>$ =$4$ + $\phi^3$ = $4.236...$) long enough to favor an
 emergence of a
 conscious life. This will continue until the whole ensemble of
 infinite quanta (bubbles,universes) begins its  collective slow
 evolution towards  the lower  ( true vacuum ) stable state of
 average dimension $\phi^3 = 0.236...$ (or $ 2+\phi^3$ if
 considered relative to the $D_o = -2$) of  extremely cold
 temperatures transforming itself into a quasi-crystal state. The
 phase transition to the lower dimensions could be viewed as the
 "big crunch" which transforms the universe into a "pancake". The
 latter is consistent with the observational data \cite{b1}.

 \smallskip
 The disruption of the metastable state is caused by vacuum
 fluctuations. These
 fluctuations can be arbitrarily large due to the fact that the
 ensemble of universes comprises an infinity of quanta (bubbles of
 different dimensions and radii). An analogous situation arises in
 the field theory where quantum fluctuations can become large since
 the respective systems have an infinite number of degrees of
 freedom. \footnote{This picture is not so unusual in physical
applications if we recall that for example Van der Waals forces
arise as a
 result of molecular fluctuations. The idea of an analogous origin
 of gravity was expressed independently  by Feynman and by Vigier
 and Petroni \cite{jv1}. Our view is that classical spacetime and
 gravity emerged from vacuum fluctuations of the Cantorian fractal
 geometry ( dimensions, for example ). Within the framework of a
 new relativity and Cantorian fractal spacetime, one now needs to
 define the " field " (which we call  Cantorian \cite{en3}) whose
 fluctuations cause the respective fluctuations of dimensions.
 Quite recently Sidharth  \cite{bs1} has also pointed out the
 importance of fluctuations in Cantorian fractal spacetime. It is
 these fluctuations that generate classical spacetime, gravity,
 and all fundamental forces of nature.} The quasi-crystal will
 eventually reach an enormous size  while evolving towards the
 regime of the minimal energy density configuration. In fact this
 follows directly form eq.(4.7) which tells us that the final
 value of the cosmological " constant " will be dramatically
 smaller than it was at the birth of the universe. This regime (
 if it will be reached in a final time interval) will indicate the
 " end " of spacetime, matter, energy,...and life.

 \smallskip
 Still we should not despair since it is quite possible that the
 cold quasi-crystallized ensemble will " collide "( in  C-space,
 outside spacetime ) with another ensemble of much higher vacuum
 energy, higher temperatures ( we are assuming a " multi-verse"),
 and the world will again begin to reheat  and climb up the
 dimensional ladder, evolving towards the metastable state of $ D =
 4 + \phi^3$. Vacuum fluctuations will trigger another life cycle
 of the emerging universe ( in general different from the one that
 descended into quasi-crystal state), and the process will repeat
 itself time and time again. A more rigorous picture of this
 cyclical universes scenario requires the construction of the
 quantum field theory in noncommutative C-spaces
 \cite{cc1,cc3,cc6}.

 \smallskip
 We can speculate that this  " cosmic dance " has been going on
 forever, and it will continue on forever within a  cyclical
 scaling time associated with the renormalization group flow. The
 cyclical renormalization group  scaling time is the true
 universal arrow of time. Its existence is predicated on the
 existence of negative  dimensions ( a new relativity sea of
 negative dimensions). On the other hand, the time measured by our
 clocks is just one of the coordinates which is interchangeable with any
 other coordinate , the fact following from the diffeomorphic
 invariance of general relativity.

 \smallskip
 Our arguments showing that there is " no cosmological constant
 problem " in a new relativity agree with Nottale's arguments who
 explained  in simple terms a huge discrepancy between the
 cosmological " constant " measured at the cosmological scales,
 with the cosmological " constant " measured at Planck scales. It
 was necessary to formulate scale relativity to point out an
 existence of a dependence of measurements' scales on a frame of
 reference. In a new relativity theory, there is no such thing as
 the cosmological " constant problem "
 because all the constants in nature are subject to
 renormalization group flow with scaling time. It is precisely
 because of a process of self-organization and self-tuning that all the
 constants in nature properly adjust themselves to the
 renormalization group scaling flow , or arrow of time. Such arrow
 of time itself materialized only because the world emerged from a
 strongly non-equilibrium state.

 \bigskip

\section{ Exact Evaluation of the Nottale's Upper Scale In Nature }

 \bigskip

 Here we determine the exact value of the minimum non-zero
 temperature of the quasi-crystal phase. We will see that is
 directly related to the maximum upper scale in nature which ( as
 we already indicated) is dual  to the Planck minimum scale.
 Nottale already gave estimates of what this scale should be. However
 at that time he was not aware of the power of the duality principle
 provided by the string theory and therefore had no way of knowing how
 to
 calculate the maximum upper scale from the basic principles. On
 the other hand, we will utilize duality and provide such an
 evaluation.

 This duality between the large and the small is at the heart of the UV/IR
 entanglement of quantum field theory in noncommutative spaces  .
 If the upper scale were infinity, and the lowest scale were zero,
 such an entanglement wouldn't be possible. The spacetime
 coordinates of a noncommutative space do not commute because the
 Planck scale is not zero,i.e $[X^i, X^j] \sim \Lambda^2 $. This
 property represents one of natural consequences of a new relativity
 : ordinary " point " coordinates $x^\mu$ are to be replaced by
 Clifford-algebra valued {\bf X} multi-vectors ( matrices ). It is
 clear that in general the latter do not commute.

 \smallskip
 As we have already argued, the Planck scale is the ultimate UV
 regulator and the upper scale is the ultimate IR regulator. If in
 preparation of the ensemble of hyper-spheres we take into account
 Nottale's scale relativistic corrections, the fractalization of
 these hyper-spheres would be taken into account from the very
 beginning. The respective  volumes will be not only radius and
 dimension dependent but also resolution dependent. We will show
 that by including Nottale's scale relativistic corrections (that
 is resolution-dependence of physical quantities) the system's
 average dimension will change from the " metastable
 "quasi-ergodic value of $<D>$ = $4$ + $\phi^3$ to the true vacuum
 value of $<D> = \phi^3$.

 \smallskip
 In Section 3 we found that the average dimensions calculated
 with the use of either the discrete sum or the integral eq.(3.12)
 is very close to the exact average dimension $d_c^{(4)}$=$4$ + $\phi^3$
 found on the basis of fractal dimensions and thus signaling the
 presence of a quasi-ergodic, metastable state. For this
 reason the world  remains for a very long period (conducive to an
 emergence and existence of life forms) in a state whose average
 topological dimension is indeed very close to the value of $ D = 4$ associated
 with a smooth spheres ( manifolds).

 \smallskip
 To precisely calculate  the average dimensions  we should include
 the scale relativistic corrections . In essence these
 corrections amount to a computation of the averages by packing
 the ensemble thermal reservoir ("box") with fractal spheres
 instead of smooth ones. The averaging requires integration with
 respect to the radii $R$ from $R_H $ to ${\cal L}$ in the
 quasi-crystal phase and from $\Lambda $ to $R_H$ in the
 metastable phase. The former case allows us to find the maximum
 upper length implicitly:

 $$I= { \int^\infty_0dD~D^2  {\sqrt \pi}^D  [ \Gamma ({ D+2 \over
 2 }) ]^{-1}  \int^{({\cal L}/\Lambda)}_ {({R_H/\Lambda})} d({R
 \over \Lambda})  ~ ({R \over \Lambda})^{D-1}~ [e^{
 {E_{min}(R/\Lambda)^D\over KT_o }} -1 ]^{-1 } J \over \int^\infty
 _0 dD~D   {\sqrt \pi}^D [ \Gamma ({ D+2 \over 2 })]^{-1} \int^{(
 {\cal L}/\Lambda)}_ {R_H/\Lambda} d({R \over\Lambda})~ ({R
 \over\Lambda})^{D-1} ~[e^{ {E_{min}(R/\Lambda)^D\over KT_o }}  -1
 ]^{ -1 }J }. \eqno (5.1)$$ where $J = \int_1^{R/\Lambda} d ({\xi
 \over \Lambda} )~ ( { \xi \over\Lambda } )^{-D (\gamma -1)}$,
 $\gamma(\xi)=(1-\beta^2)^{-1/2}$ and
 $\beta^2={ln^2(\xi/R)}/{ln^2(R/\Lambda)}$ are Nottale's
 scale-relativistic parameters, and $\xi$ is the resolution
 $\Lambda \leq \xi \leq R$. At the ultimate resolution allowed in
 nature $\xi=\Lambda$, and the volume\footnote
 {$D[\gamma(\xi)-1]=D_F-D_T$ is the difference between the fractal
 (Hausdorf) and topological dimensions} of a fractal hyper-sphere
 $V=V_0\xi^{-D[\gamma(\xi)-1]}$ goes to 0 meaning that the
 infinite-dimensional limit is reached where spacetime evaporates
 into a sea of fractal dust, the hyper-point. This would require
 an infinite amount of energy.

 \smallskip
 Generally speaking, the average dimension given by eq.(5.1)
 depends on two parameters $a= E_{min} / KT_o$ and $b = {\cal L} /
 \Lambda $, whose values should be chosen in such a way as to fit
 the experimental observations. Thus effectively our theory is
 a 2-parameter theory which is a great advantage as compared with
 other theories involving much more phenomenological parameters.

 For example one can choose $a=b$ as a test condition. This yields
 an integral equation that will define the maximum upper scale $
 {\cal L}$ exactly when we require the average dimension ( with
 respect to the zero reference point  $D_o = -2 $ ) to be :

 $$ <D' > = 2+ \phi^3 = 2.236....= I_1(E_{min}/ KT_o  = {\cal
 L} / \Lambda~; {\cal L} / \Lambda;~  R_H/\Lambda). \eqno
 (5.2) $$ where $I_1$ is the value of I for $a=b$. Result (5.2) is
 tantamount to choosing an average value of $<D> = \phi^3$. This
 value reflects a decrease (on average) by one of each of the three
 spatial and one temporal dimensions ( plus fluctuations ) of the
 metastable vacuum whose overall average dimension is $<D>$ = $4$ + $\phi^3$.

 \smallskip
 In particular, scale relativistic corrections can shift the value
 $ < D' > = 6.38....$ to the  exact  value of $6$+$\phi^3$ = $6.236 $
 by carefully selecting the value of the Hubble radius $R_H$ at the end
 of the metastable phase  before it begins its evolution
 towards the true vacuum , the quasi-crystal phase. The respective
 temperature $T_H$ ( which is less than the $T_{COBE}$ ) is found
 from the following expression:

 $$ { E_{vacuum }\over K T_{H} } = {R_H \over \Lambda } . \eqno
 (5.3) $$ This temperature will influence the second integrals in
 (5.1),namely the integrals involving the Bose-Einstein distributions.
 Including the scale relativistic corrections, and integrating
 between the limits $\Lambda $ and $R_H$ we  arrive at one more
 integral equation:

 $$6   \epsilon = I_2(E_{vac} /KT_H = R_H/ \Lambda ~; R_H/
 \Lambda)= 6+\phi^3 = 6.236....\eqno (5.4) $$ Equation (5.4)
 will give the exact ratios of $R_H/ \Lambda = E_{vac}
 /KT_{H}$ to fit the metastable fixed renormalization group point(
 relative to the zero point $D_o = -2 $) of $6 + \phi^3$ . Here
 $I_2$ is the value of $I$ for $E_{vac} /KT_H = R_H/ \Lambda$.

 It is seen that without using fractal spheres and the respective
 Nottale's scale relativistic corrections it is impossible to fit
 the metastable state's value $ 6 + \phi^3 $  . The latter is
 calculated relative to  $ D_o = -2 $ which implies that $<D> =
 4+\phi^3$. Such value precisely corresponds  to the fractal
 dimensions of a set structure living "underneath" a  four
 dimensional smooth sphere. $ d_c^{(4)}$ =$4$ + $\phi^3$ is exactly
 the Cantorian fractal dimension of the set ${\cal E} ^{ (4)}$.

 Thus we have demonstrated an existence of  a very close
 connection of the Cantorian fractal spacetime model and Nottale's
 scale relativity. Moreover,
 using duality arguments of the string theory  we were able to
 exactly define the upper Nottale's scale ${\cal L} $  via the
 integral equation that requires fitting $\phi^3$ as the true
 dimension of the true vacuum of our theory.

 It is tempting to speculate that because there are $3$ different
 scales, $\Lambda, R_H, {\cal L } $ they are not necessarily
 independent. As a test function connecting these scales we can use
 the mean geometric formula:

 $$ R^2_H = {\cal L }\times \Lambda ~\Rightarrow (R_H/\Lambda ) = (
 {\cal L} / R_H ) ~ \Rightarrow    ({\cal L } / \Lambda) =
 (R_H/\Lambda )^2.  . \eqno (5.5) $$

 The final verdict on the validity of (5.5) can be pronounced only
 if we will be able to calculate the integrals of (5.1).
 \bigskip

 \section{Conclusion}
 \bigskip

 It is our thesis that the evolutionary process of the universe is
 closely connected to the dynamics of self-organized
 criticality, complex systems, self- referential noise and quantum
 dissipative processes as fundamental aspects of reality. The
 universe as it evolves simultaneously self-tunes. Roughly
 speaking, this process could be described by what Finkelstein has
 called a  variable quantum law, or $q$-process.

 Clearly the naive Lagrangian formalism used in ordinary spaces will
 not work here. The master action functional  \cite{aa1,cc2,cc3}

 $$ S\{\Psi [X(\Sigma)]\}= \int [{1\over 2} ({ \delta \Psi \over
 \delta X }*{ \delta \Psi \over \delta X }+ {\cal E}^2 \Psi*\Psi)+
 {g_3\over 3!} \Psi*\Psi*\Psi + {g_4\over 4!}
 \Psi*\Psi*\Psi*\Psi]DX(\Sigma).\eqno (6.1) $$

 of quantum field theory in C-space "lives" outside spacetime and
 as such requires a more general formalism. The latter is provided
 by the renormalization group approach which is an essential
 ingredient of the propagation of strings in curved spacetime
 backgrounds. The curved spacetime itself is  the solution of the
 coupled Einstein-Yang-Mills equations which can be described as
 the vanishing of the beta functions associated with the world
 sheet couplings. The $scaling$ dynamics encodes the $motion$
 dynamics of the strings.

 \smallskip
 The quantum group symmetry of such noncommutative quantum field
 theory is described by Braided Hopf Quantum Clifford algebras.
 The respective vertices are described as follows. The 2-point
 vertex corresponds to the pairing of the quantum algebra. The
 3-point vertex is given by product and coproduct of the quantum
 algebra, that is annihilation of two C-lines and generation of
 the third line, or creation of two C-lines out of one C-line.

 The Cantorian field $\Psi[X(\Sigma)]$ is a hyper-complex number
 of Clifford algebra-valued object. In particular it could be
 quaternionic or octonionic-valued. The C-lines in C-space are
 nothing more than the Clifford-algebraic extension of Penrose's
 twistors into a complex field. Action (6.1) is unique in a sense
 that the  braided Hopf quantum Clifford algebra fixes the types
 of terms allowed by the action. Quantum fluctuations of the field
 $\Psi[X(\Sigma)]$ of the Cantorian fractal Spacetime were
 responsible for the creation of the present quantum universe.

 \smallskip
 In $D = \infty$ limit it may be possible to construct a unique
 topological action for the world. The large $D=N$ limit was
 discussed briefly in \cite{cc9} with respect to the
 relations between conformally invariant $\sigma$ models on anti
 de Sitter spaces, $AdS_{2N}$, and Zaikov's Chern-Simons $p'$-brane
 field theories residing on the projective boundaries of anti de
 Sitter spaces, $S_{2N-1}$ spheres. When $D=2N$ and $2N=p'$ tend to
 $\infty$ then there is no distinction between $D$ and $p'$.
 As  a result, Zaikov's Chern-Simons $p'$-brane quantum field theory
 is the natural candidate for the topological field theory for the
 world. In this case ($2N=D\rightarrow \infty$) the Chern-Simons
 $p'$-brane becomes the infinite-dimensional spacetime filling
 $p$-brane \cite {cc2}.

 \smallskip

 Zaikov's Chern-Simons classical $p-brane$ field theory admits
 $W_{{1\infty} }$ algebras as an algebra of constraints. Its respective
 connection to Vasiliev's higher spin conformal field theories
 \cite{v1}, $W$-geometry , and $W_{\infty}$ algebras based on
 Moyal-Fedosov quantization was given in \cite {cc10}. In the
 limit $D \rightarrow \infty$ one encounters a transition to the
 transfinite continuum of Cantorian fractal spacetime. This would
 require a study of infinite-dimensional loop spaces and
 associated loop algebras.

 \smallskip
 Finally, we believe that the evolving universe constructs its own
 Hilbert space. Hence,  nonlinear complex dynamical systems  has
 to be an essential part of reality. String theory, noncommutative
 geometry, quantum groups, Hopf algebras and the new scale
 relativity, for example, have  already shown that at the
 fundamental level coordinates do not commute, and the Heisenberg
 uncertainty relations are to be modified to account for the Planck
 scale to be the minimum attainable length in nature. As we try to
 compress the strings (membranes, $p$-branes...) to scales smaller
 than the Planck scale the strings (membranes, $p$-branes...) begin
 to grow in size signaling that there is a nontrivial ultra
 violet/infrared entanglement. Let us not forget that the Planck
 scale should not to be confused with the string scale $l_s$  since
 for example the $D$-branes can probe distances smaller than $l_s$.
 Non-Archimedean geometry and $p$-Adic numbers are the natural
 geometry and numbers required by the New Relativity at Planck
 scales consistent with the Cantorian-Fractal nature of spacetime
 \cite{matt}.

 \smallskip
 Self-tuning of the universe in process of its evolution implies
 that values of the fundamental "constants", including the
 cosmological constant are also adjusted accordingly. In
 particular the classical $spacetime$ of our perception is
 constantly evolving entity which does not exist $ab~initio$. We
 can say that it is a $process$ in the making \cite{rc1}. Thus we
 can abandon the idea of a single universe  inflated to the size
 observed today. Instead we can introduce a true q-process
 (self-referential)  with a hierarchical family of universes
 (reminiscent of  a  matryoshka doll) where each representative
 member of the family has an  average dimension (as seen today) of
 approximately $4$ + $\phi^3$ = $4.236067977...$. Therefore within this
 model we view our world only as a representative of an infinite
 {\bf ensemble } of universes instead of a "given " and fixed
 universe inflated to the sizes of today's universe. This ensemble
 approach allowed us to calculate universe's average dimension by
 using the ensemble averaging. If we look at this differently, we
 in fact performed the Feynman path integration over the infinite
 possible scenarios/histories of the world.

 \smallskip
 A world as we see it is a "perceptual averaging projection" of  a
 perpetually changing processes underlying the visible world of
 our senses. The world tomorrow will not the same as the world
 today. Quoting Wheeler, it is possible to say that we ourselves
 are true observers and participants in this averaging process
 which we  perceive as reality. This represents (in
 different terms)  the Everret-De Witt-Wheeler many worlds
 interpretation of quantum mechanics. We perform billions of
 measurements every day using one of the most sophisticated
 measuring devises : our brains \cite{dc1}.  Every time an
 observation is made, a branching of possible scenarios will
 occur. For this reason  we wholeheartedly subscribe to Penrose's view
 that the physics of the human brain ought to be included into
 future physical theories \cite{dc1,dko1}.

 \bigskip

 \centerline{\bf Acknowledgements }
 \smallskip

 We thank D. Chakalov, S.Paul King , D. Finkelstein, E. Spallucci,
 E. Guendelmann , G. Chapline G. Bekkum and R. Guevara for many
 discussions.   One of us (CC) wishes to thank A. Cabo, H. Perez, C. Trallero, D.
 Villarroel, M. Chaichan. To C. Handy. M. Handy, D. Bessis, A.
 Boedo, M. Bowers, A. Bowers our deepest gratitude for their
 assistance and encouragement.

 \bigskip

 \section{Appendix A}

 \smallskip
Using the notion of a quantum "path" as a fractal curve
\cite{rf1},\cite{ab1} with poles at the time interval $\Delta t
=0$ and shifting this value to the minimum attainable time
interval , Planck's time $\tau_P$

 $$ \Delta t = \tau_P \eqno(A1)$$

we get the following (cf.\cite{rf1})

 $$E =\frac{m}{2}{(\frac{\Delta x}{\Delta t})}^2 =
 \frac{a}{\Delta t - \tau_p}. \eqno(A2)$$

 where a is a constant to be determined. We rewrite (A2)
 $$\frac{m}{2}{(\frac{\Delta x}{\Delta t})}^2=
 \frac{m}{2}{(\frac{\Delta x}{\Delta t})}^2\frac{\tau_p}{\Delta
 t}+\frac{a}{\Delta t}. \eqno(A3)$$
 If we divide both sides of (A3) by
 $m \Delta x/ 2({\Delta t})^2$ and introduce the average momentum
 $<p>$= $m(\Delta x/\Delta t)$ then (A3) yields a relation between
the spatial resolution $\Delta x$ and the average momentum:
$$\Delta x = \tau_P {<p>\over m}  + \frac{2a}{<p>}. \eqno(A4)$$

 Since by operational definition the resolution of a physical
 device $\Delta x$ must be less ( or equal) than a statistical
 mean square deviation $\Delta x_s$we get from (A4)

$$\Delta x_s \geq \Delta x=\tau_P {<p>\over m} + \frac{2a}{<p>}.
\eqno(A5)$$

 The minimum of the r.h.s. of (A5) is reached at

 $$\Delta x_{min} =2\sqrt{2\tau_P{a\over m}}\eqno(A6)$$

which in particular means $\Delta x_s \geq \Delta x_{min}.$ Since
the average momentum $<p>$ is  of the order of the mean square
deviation,that is $<p>^2 \sim (\Delta p_s)^2$ we get from (A6) the
following

 $$\Delta x_s\Delta p_s \geq {\tau_P\over m}{(\Delta p_s)}^2+ 2a.
 \eqno(A7)$$ If we set $\tau_p$ =$0$ we recover from (A7) the conventional
 uncertainty relation of quantum mechanics which means that $2a
 =\hbar$.  By setting the min resolution $\Delta x_{min}$ =
 $\Lambda$ we obtain from (A6) that

 $${\tau_P\over m}=\frac{\Lambda^2}{4\hbar}.\eqno(A8)$$

 Upon inserting this value and $2a=\hbar$ back into (A7) we arrive
 at the following

$$\Delta x_s \Delta p_s \geq
 \hbar[1+\frac{1}{2*2!}{(\frac{\Lambda}{\hbar})}^2 {\Delta
 p_s}^2]. \eqno(A9)$$

which is a truncated stringy uncertainty relation of string
theory. The right hand side of (A9) (with $ <\delta p_s>
 \sim <p>^2=\hbar^2 k^2$) represents an "effective" Planck constant
 $$h_{eff}=\hbar[ 1+ \frac{1}{2*2!}\Lambda^2k^2].\eqno(A10)$$

 \section{Appendix B}

 \smallskip

 { \bf  An Elementary Derivation of the Area-Entropy Relation In
 Any Dimension}

 In this appendix we review the derivation of the area-entropy
 relation and explain why the stringy holographic principle is a
 direct result of a new relativity. Let us consider an
 infinite-dimensional quantum spacetime quantized in discrete
 geometric bits of point, {\bf 1}-loop, {\bf 2}-loop ,...,
 $p$-loop  histories. The latter play the same role as photons in
 quantum electrodynamics with a difference that $p$-loop histories
 ( quanta of spacetime) form a self-referential ( bootstrapping)
 medium of a dynamical spacetime in the making. Thus counting the
 number of these quanta we will get the information content of the
 attendant spacetime.

 \smallskip
 It has turned out that the appropriate counting mechanism is
 provided by Clifford algebra. In fact, a Clifford algebra in $D$
 dimensions or degree $D$ has $2^D$  independent " components "
 that represent the total number of the point , holographic area,
 holographic volume, holographic four volume,etc. coordinates
 associated with the hierarchy of point, {\bf 1}-loop, {\bf
 2}-loop, ... histories, or excitations of $D$-dimensional
 spacetime \cite{cc1}-\cite{cc4} . This is simultaneously the
 total number of  "geometric  quanta" $N$, meaning
 that $N$=$2^D$. The respective Shannon information
 entropy is

 $$S = K ln~ 2^D = K D ln~2.   \eqno (B1) $$
 \begin{figure}
 \begin{center}
 \includegraphics[width=6cm, height=6cm]{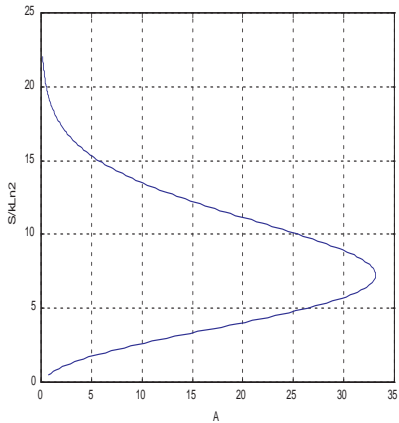}
 \caption{\small Graph of a dependence entropy vs. area for a unit
 hyper-spheres}
 \end{center}
 \end{figure}
 Therefore $$ D = { S \over K ln~2 }\equiv s.\eqno(B2)$$ On the
 other hand,the $(D-1)$-dim  hyper-area enclosing a $D$-
 dimensional( bulk) hyper-volume  of a $D$-dimensional  sphere of
 a  unit Planck radius is :

 $$ A_{D-1} = D{ {\sqrt \pi}^D \over \Gamma( { D+2 \over 2 } ) } .
 \eqno (B3)  $$ which means that $D$ and hence $s$ ( eq.B1) is an
 implicit function of the  hyper-area $A$: $$ s = D(A)\eqno(B4)$$
 Since this function cannot be expressed analytically we restrict
 ourselves to its graphical representation shown in Fig. 1. It is
 easily seen that the graph has a region of a linear dependence
 $s$ vs. $A$ in a rather narrow range of dimensions, namely from $
 D=2 $ to $D=5$.

 \smallskip
 The slope of the linear region is $1/6.9$ which is less than the
 Bekenstein upper bound. For the dimensions outside this range the
 dependence clearly deviates from linearity. If we consider a
 sphere with a non-unit radius then a respective graph of $s$ vs.
 $A$ will look similar to Fig.1, with the only difference that for
 the same $s$ the values of the respective areas $A$ will increase
 with the radius increase, as in Fig 2.

 Let us apply this result to macroscopic black holes.We can
 consider a macroscopic black-hole as being built from many "bits"
 or mini black holes of unit Planck radius. The state of a
 macroscopic black-hole is evaluated by simply "counting " the
 number of micro-states accessible to the macroscopic system.

 Since Bekenstein and Hawking assumed that there exists a linear
 relation between area and entropy (valid only for a certain range
 of dimensions, as was shown above) , the entropy could be found by
 simply adding elementary portions (or bits) of areas. In other
 words, a linear superposition of the fundamental mini black holes
 states was possible due to the  linearity of the area-entropy
 relation . A more general way of combining bits of
 information-entropy is as follows.
 \begin{figure}
 \begin{center}
 \includegraphics[width=6cm, height=6cm]{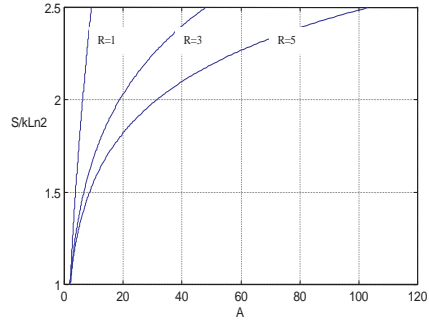}
 \caption{\small Graph of a dependence entropy vs. area for
 hyper-spheres of 3 different radii: R=1, 3, and 5}
 \end{center}
 \end{figure}

 \smallskip
 Suppose we wish to calculate the entropy of a  cubic object.
 Since a cube is topologically equivalent to a sphere, we simply
 deform
 the cube into a sphere, simultaneously  preserving the cube's area.
 This  leads to an increase of the sphere's volume as compared
 to the original cube. As a next step we evaluate the surface area of
 the sphere  suitably normalized to Planck units. If now we divide
 this area
 by the area of a unit sphere (in Planck units) then the result would
 give us
 the information entropy ( divided by $KLn2$) of the original cube.

 \smallskip
 However there are two obvious objections to this procedure:

 1.Is it possible to pack all  unit spheres into the
 big sphere without leaving any voids ?

 2. What if the information-entropy, given by the ratio of areas
 (in units of $kln~2$) is not an integer ?

 To answer the first objection we perform an inverse deformation
 of each unit sphere into a fundamental small cube of the same
 area, then we deform the large sphere  into the original cube of
 the same area. Now we can  pack all the elementary cubes into a
 large cube without leaving voids.

 The second objection is a little bit more tricky. In this case a
 small region of the cube remains unpacked. To resolve this
 problem we use properties of fractal geometry: the large cube is
 to be filled with cubes of fractal dimensions. Since fractals
 have a property of space filling the cube will be filled without
 any voids left.

 Now we make a plausible assumption that only one bit of
 information can be ascribed to a minimum attainable scale, the
 Planck scale $\Lambda.$ This assumption is supported by the
 holographic origins of  chaos in $M$  theory \cite{ia1}. An onset
 of chaos is signaled by impossibility to pack the energy levels (
 spectra ) into small regions of very high energy ( spectral )
 density distributions. The repulsive ( holographic ) feature of
 these spectral lines is the indication of quantum chaos.

 This means that for every two adjacent cubes/cells packed side by
 side, information will be "pushed" out to the respective
 surfaces. As a result their common boundary surface would contain
 two bits of  information  per unit area (in Planck units) , which
 contradicts our initial assumption that  one cannot have more
 than one bit per unit  area ( an " exclusion principle ").
 Applying this  argument to the overall packing-process  we arrive
 at a situation analogous to the Gauss's law in electrostatics: the
 information can reside only on the boundary surface of a cube, and
 hence on the surface of  the original sphere. For the importance
 of the Cantorian fractal spacetime model for the theory of
 superconductivity, and its potential high temperature
 applications, see \cite{Agop}.

 \smallskip
 In general, the graphs $s = D(A)$ (figs. $1$ and $2$) exhibit a
 variable
 slope indicating that information density changes with the area.
 The slope reaches infinity in two cases: 1) at a maximum
 attainable area and 2)in the limit of $D \rightarrow \infty$ when
 the area goes to $0$. The latter signals an  onset of a phase
 transition under which spacetime "evaporates" leaving behind a fractal
 Cantor dust. Since the area tends to $0$ the information
 density blows up.

 \smallskip
 In the former case ( for simplicity consider a
 unit sphere) the slope becomes infinite between $D=4$ and $D =5$.
 This is the metastable renormalization group fixed point of
 $D$= $4$+$\epsilon$ = $4$ + $\phi^3$. Not surprisingly, this is another
physical reason why we live in $D=4$. Can we actually reach $D
=\infty $ ? In other words, how much energy this
 process would require? In \cite{cc4} we have discussed why
 dimensions,
 energy, information  are indistinguishable at the ultimate scale
 provided
 by nature, the Planck scale $\Lambda$. Once Nottale's scale
 relativistic
 corrections to  resolutions of physical devices are  taken into
 account,
 it turns out that to reach $D =\infty$ (which is tantamount to
reaching the Planck scale \cite{cc5}) one needs an infinite amount
of energy.

 \section{Appendix C}
 {\bf Renormalization group flow and a local scaling arrow of
 time.}

 \smallskip
 Here we would like to point out a connection of our work to the
 renormalization group techniques of quantum field theory within
 the context of fractals. The emergence of a microscopic Liouville arrow
 of time within the context of $D$-branes and the breakdown of
 Lorentz invariance at very short scales has been analyzed by
 Ellis et al \cite{ellis}. In  a sense, a  picture of an infinite
 unfolding of a hierarchy of infinite-dimensional worlds
 represents a visualization of actions of the renormalization
 group. In simple terms we can recreate this picture as follows.
 Let us focus our eyes at a small region of a monitor which we will
 conditionally call a "point". Certainly one can even imagine to
 be "inside" this point. A physical realization of this immersion
 would require a lot of energy which will increase with a decrease
 of the point's size. As more energy is supplied to get inside
 the "point " (corresponding to a greater zoom) an observer
 will notice unfolding of additional dimensions. This resembles
 playing with a toy matryoshka doll. Since energy
 generates information ( and generated by information) an increase
 of energy in this process will generate additional information
 and as a result will reverse a  local microscopic scaling arrow of
 time.

 We begin to move into a different hierarchical family of the
 Cantorian matryoshka-like spacetimes whose local average
 dimension will be greater than our initial $D = 4$. The opening
 worlds would be the members of the family which belongs to our
 "past" with respect to the local scaling renormalization group
 arrow of time. However to reverse the renormalization group arrow
 of time of the whole (global) universe is a totally different
 enterprise because energy involved in achieving such a feat would
 be enormous, of the order of the total mass-energy of the whole
 universe.

 Kreimer \cite{dk1} has provided an ample proof that the
 renormalization and combinatorial techniques of quantum field
 theory contains a Hopf algebra associated with the Feynamn's
 diagrams. A multi fractal, multi-vector , and multi-scaling basis
 of  a new relativity allows one to utilize a rich and diverse
 mathematical apparatus  of quantum groups, noncommutative
 geometry, Feynman graphs, Hopf algebras, etc. as an
 algebraic-combinatorial and self-recursive way of coding the
 infinite unfolding of the transfinite  Cantorian multi fractal
 hierarchical matryoshka doll-like worlds.
 Interestingly enough, the Euler gamma function ( which is
 self-recursive and which plays an important role in our
 derivation)is an essential tool in the renormalization process of
 quantum field theory.

 To make our statement about reversing a scaling arrow of time
 more precise we invoke Zamolodchikov's central charge theorem
 \cite{az1}  which originally was formulated for flows in conformal
 field theories in $D=2$ but can be easily generalized to $D=2n$.
 The theorem states that the central charge $c[g^i, \mu  ]$
 monotonically decreases along its renormalization group trajectory
 moving towards the infrared region

 $${ \partial g^i \over \partial ln ~\mu }~ { \partial c [g^i, \mu
 ]\over  \partial g^i} = { \partial c [g^i, \mu ]\over  \partial
 ln~\mu } \le 0. \eqno (C1)$$
 Here $g^i$ are the world sheet couplings and $\mu$ is the
 scale,or the subtraction point.

 On the other hand, in string theory and conformal field theory
 central charges are  directly related to dimensions. For example,
 critical strings are anomaly free in $D=26$ to cancel
 the $c = -26$ charge associated with the ghosts; critical
 superstrings are anomaly free in $D=10$, etc.  This fact
 indicates that our assertion about " reversal" of  the scaling
 renormalization group flow on its way to regions of higher
 central charges and hence higher dimensions is compatible with
 Zamolodchikov's theorem. Moreover, there exists a connection
 between conformal field theory , the monster group, and sphere
 packing which could be found in ref. \cite{gn1}. Roughly
 speaking , the monster group associated with the Golay error
 correcting code for a self-recursive system encodes the
 unfolding process into higher dimensions. And conversely, upon
 reversal of  the scaling arrow of time, the monster group encodes
 condensation of dimensions from the higher( on the average) initial
 values to the smaller ( on the average ) values. This could be viewed
 as coding a monotonic (quasi) crystallization process. Similar
 ideas about a code underlying spacetime being a {\bf q-network} of
 {\bf q-processes} was expressed by Finkelstein long time ago
 \cite{df1}.

 \end{document}